\documentclass[conference]{IEEEtran}
\usepackage{subcaption}
\usepackage{booktabs} 
\usepackage[normalem]{ulem}
\IEEEoverridecommandlockouts
\usepackage{cite}
\usepackage{amsmath,amssymb,amsfonts, bm, bbm}
\usepackage{algorithmic}
\usepackage{graphicx}
\usepackage{textcomp}
\usepackage{xcolor}
\DeclareMathOperator*{\argmin}{arg\,min}
\DeclareMathOperator*{\argmax}{arg\,max}
\def\BibTeX{{\rm B\kern-.05em{\sc i\kern-.025em b}\kern-.08em
    T\kern-.1667em\lower.7ex\hbox{E}\kern-.125emX}}
\begin{document}

\title{Patch-based learning of  adaptive Total Variation parameter maps for blind image denoising\\
\thanks{$^*$ This author carried out this project in the joint Inria and CNRS MORPHEME team, UMR7271, Laboratoire I3S, Sophia-Antipolis, France.

The authors acknowledge the support received by the ANR-21-CE48-0008 PRC MICROBLIND project and of the ANR-22-CE48-0010 JCJC project. LC acknowledges the financial support of the European Research Council (ERC StG grant MALIN, 101117133).}
}

\author{\IEEEauthorblockN{Claudio Fantasia$^*$}
\IEEEauthorblockA{\textit{Politecnico di Torino} \\
\textit{DAUIN}\\
Turin, Italy \\
claudio.fantasia@polito.it}
\and
\IEEEauthorblockN{
Luca Calatroni$^*$}
\IEEEauthorblockA{\textit{MaLGa Center, DIBRIS,} \\
\textit{Università di Genova},\\
 \textit{MMS, Istituto Italiano di Tecnologia}, \\
Genoa, Italy \\
luca.calatroni@unige.it}
\and
\IEEEauthorblockN{Xavier Descombes}
\IEEEauthorblockA{\textit{Université Cote d’Azur} \\
\textit{INRIA, CNRS, I3S}\\
Sophia Antipolis, France \\
xavier.descombes@inria.fr}
\and
\IEEEauthorblockN{Rim Rekik}
\IEEEauthorblockA{\textit{Université Grenoble-Alpes} \\
\textit{INRIA, CNRS}\\
Grenoble, France \\
rim.rekik-dit-nekhili@inria.fr}
}

\maketitle

\begin{abstract}
We consider a patch-based learning approach defined in terms of neural networks to estimate spatially adaptive regularisation parameter maps for image denoising with weighted Total Variation (TV) and test it to situations when the noise distribution is unknown. As an example, we consider situations where noise could be either Gaussian or Poisson and perform preliminary model selection by a standard binary classification network. Then, we define a patch-based approach where at each image pixel an optimal weighting between TV regularisation and the corresponding data fidelity is learned in a supervised way using reference natural image patches upon optimisation of SSIM and in a sliding window fashion. Extensive numerical results are reported for both noise models, showing significant improvement w.r.t.~results obtained by means of optimal scalar regularisation. 
\end{abstract}


\section{Introduction}
Hybrid approaches in the field of image restoration and image reconstruction represent nowadays a popular compromise between the good theoretical and interpretability properties of classical model-based approaches and the effectiveness of deep-learning data-driven procedures. Prominent examples in this respect are, for instance, algorithmic unrolling \cite{monga2021algorithm} and plug-and-play procedures \cite{Kamilov2023} aiming at incorporating deeply-learned ingredients (i.e., regularisation functionals, their proximal operators) within standard regularisation schemes. \\
The task of image denoising offers an enlightening playground for the development of hybrid procedures, given the number of datasets and competing methods available nowadays. The task there consists in computing a noise-free version $\bm{x}\in\mathbb{R}^n$ from a noisy image $\bm{y}\in\mathbb{R}^n$ under some prior knowledge of the type of noising process desribed by a (possibly non-linear) function $\mathcal{T}:\mathbb{R}^n\to\mathbb{R}^n$ so that:
\begin{equation}
    \bm{y} = \mathcal{T}(\bm{x}),
    \label{eq:Intro_InverseProblem}
\end{equation}
Classical examples are, for instance, additive white Gaussian noise where $\mathcal{T}(\bm{x}) = \bm{x}+\bm{e}$ where $\mathbf{e}\sim\mathcal{N}(0,\sigma^2 \textbf{Id})$ and signal-dependent Poisson noise where $\mathcal{T}(\bm{x})=\mathcal{P}(\alpha \bm{x}+\bm{\eta})$, where $\alpha>0$, $\bm{x}\in\mathbb{R}^n_{\geq 0}$ and $\bm{\eta}\in\mathbb{R}^n_{>0}$ is used to have a strictly positive parameter.

\begin{figure}[h]
    \centering
    \begin{subfigure}[b]{0.55\linewidth}
        \centering
        \includegraphics[width=\linewidth]{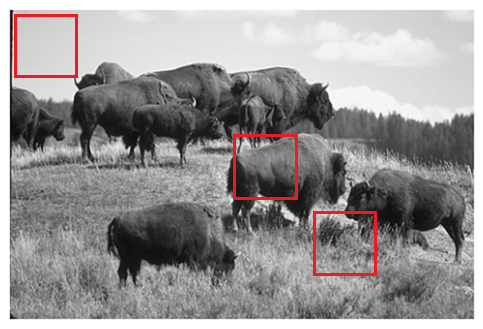}
        \caption{Clean image.}
    \end{subfigure}
    \begin{subfigure}[b]{0.35\linewidth}
        \centering
        \includegraphics[width=\linewidth]{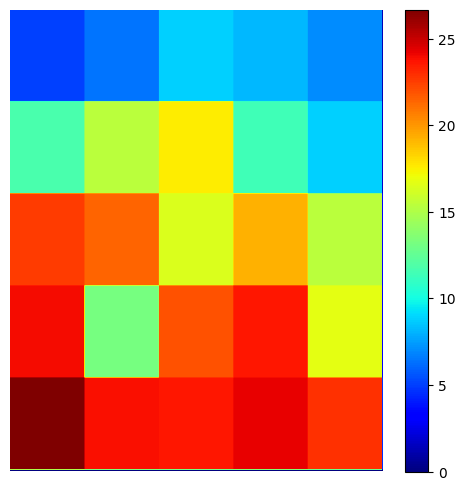}
        \caption{Optimal $\bm{\mu}_{\text{map}}$.}
    \end{subfigure}

    \begin{subfigure}[b]{0.32\linewidth}
        \centering
        \includegraphics[width=\linewidth]{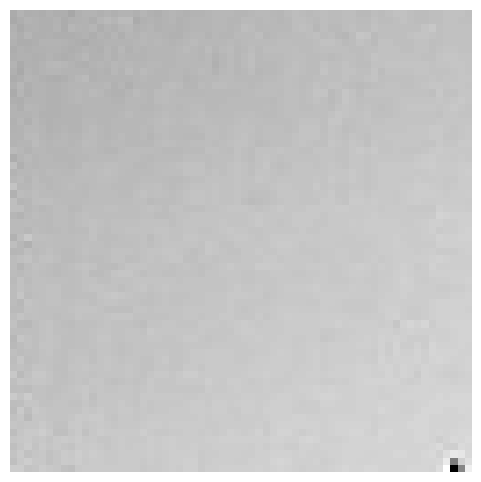}
        \caption{$\hat{\mu}_1 = 2.3$}
    \end{subfigure}
    \begin{subfigure}[b]{0.32\linewidth}
        \centering
        \includegraphics[width=\linewidth]{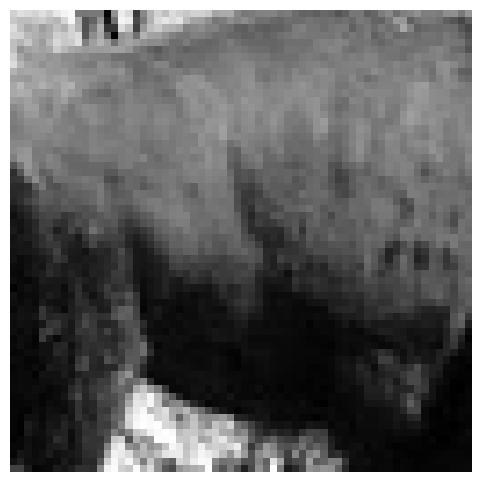}
        \caption{$\hat{\mu}_2 = 14.5$}
    \end{subfigure}
    \begin{subfigure}[b]{0.32\linewidth}
        \centering
        \includegraphics[width=\linewidth]{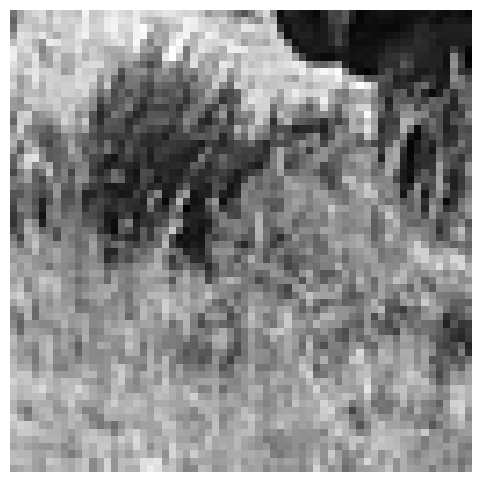}
        \caption{$\hat{\mu}_3 = 27.3$}
    \end{subfigure}

    \caption{Space-variant Gaussian denoising for Gaussian noise with  $\sigma^2 = 0.01$ with some close-ups in correspondence of some different geometric/textured regions.}
    \label{fig:numerical_reason}
\end{figure}

Following a standard Bayesian approach, a solution of \eqref{eq:Intro_InverseProblem} can be found by maximising the posterior probability density function $\pi(\bm{x|\bm{y}})$, which via the Bayes formula corresponds to solve
\begin{equation}
    \underset{\bm{x}}{\argmax} \; \pi(\bm{x}|\bm{y}) = \frac{\pi(\bm{y|\bm{x})}\pi(\bm{x})}{\pi(\bm{y})}.
\end{equation}
where $\pi(\bm{y})$ is a 
constant as $y$ is known. 
Under the assumption of negative log-concave priors and likelihood terms, 
we can equivalently consider the  problem of minimizing the composite functional being the sum of  a data fidelity $\mathcal{D}(\cdot;\mathbf{y})$ and a regularisation term $\mathcal{R}$ in the form of
\begin{equation}
     \underset{\bm{x}}{\operatorname{argmin}} \; \mu \mathcal{D}(\bm{x};\bm{y}) +  \mathcal{R}(\bm{x}),
    \label{eq:invariant formulation}
\end{equation}
where the choice of the data term and of the regulariser depend on the particular noise-distribution/a-priori information, respectively, available on the data/solution and the scalar hyper-paramter $\mu>0$ balances the effect of the two.\\
Two prominent choices corresponding to the negative logarithm of the Gaussian and Poisson log-likelihoods are the Gaussian least-square fidelity \cite{Rudin1992}:
\begin{equation}
    \mathcal{D}_G(\bm{x};\bm{y})  = \sum_{i=1}^n d_G(x_i;y_i) = 
    \sum_{i=1}^n \frac{1}{2}(x_i - y_i)^2,
    \label{eq:Gaussian_Fidelity}
\end{equation}
and the Poisson Kullback-Leibler (KL) divergence \cite{Bertero_2009} defined for $\bm{x}\geq 0$:
\begin{align}\label{eq:Poisson_Fidelity}
    & \mathcal{D}_P(\bm{x};\bm{y})  = \text{KL}_{\eta}(\bm{x};\bm{y}) \\
    & =\sum_{i=1}^{n} d_{P}(x_i;y_i)  = \sum_{i=1}^{n} y_i \log \left( \frac{y_i}{x_i + \eta} \right) + x_i - y_i, \notag
\end{align}
where we defined the 1D components $d_G$ and $d_P$ in view of the following modelling. Note that we do not consider the variance in the Gaussian log-likelihood term as it is taken into account in parameter $\mu$. 
The regularisation term $\mathcal{R}$ encodes \textit{a priori} assumptions on the desired solution (e.g. sparsity, smoothness). A celebrated choice is the following smoothed version of the Total Variation (TV) regularisation \cite{Rudin1992} which favours piece-wise constant smoothing and enjoys convexity and edge-preserving behaviour. It reads:
\begin{equation}
    \text{TV}_\epsilon(\bm{x}) := \sum_{i=1}^{n} \left\| (\bm{Dx})_i \right\|_{2,\epsilon} = \sum_{i=1}^{n} \sqrt{(\bm{D_h x})_i^2 + (\bm{D_v x})^2_i + \epsilon^2},
    \label{eq:Def_TV_norm}
\end{equation}
where $\bm{Dx} = [\bm{D_h x}; \bm{D_v x}]^T $ denotes the image gradient and where the parameter $\epsilon>0$ makes \eqref{eq:Def_TV_norm} differentiable. Optimal selection strategies for the parameter $\mu$ in the context of TV-regularised image reconstruction problems typically rely on grid-search approaches, generalised cross validation/L-curve \cite{Hansen2010} and statistical estimators exploiting the prior knowledge on noise intensity \cite{Wen2012}.\\
Traditionally, the parameter $\mu$ is a scalar value, so that regularisation is applied uniformly across the entire image domain.
More recent approaches (see, e.g., \cite{Pragliola2023}) consider weighted, that is spatially-adaptive, variants of TV-based  models where a pixel-dependent weighting between regularisation and data term is used to account for differences between textured and geometric features, thus avoiding unbalanced  effects corresponding to a \emph{global} selection of $\mu$, see Fig. 
\ref{fig:numerical_reason}, where the regularisation parameter maps is computed in a patch-wise fashion and optimal weighting values are computed on an exemplar image corrupted by Gaussian noise with $\sigma^2 = 0.01$.

The extension of the aforementioned parameter selection strategies for the computation of such pixel-dependent parameter map is often computationally unfeasible due to the dimensionality of the map (as big as the desired image), hence statistical \cite{Calatroni2019} and highly-parametrised deep-learning based \cite{Kofler2023} approaches have been proposed for its computation.

\subsection{Contribution}
We consider a patch-based learning approach for estimating adaptive, i.e.~spatially-varying regularisation maps for the problem of image denoising with unknown noise distribution. The regularisation map $\bm{\mu}_{\text{map}}$ is estimated in a supervised learning fashion where, for each noisy patch $\bm{y}_p$, the map $\bm{y}_p\mapsto \mu_i$ providing an estimate of the regularisation parameter to-be-located in the pixel $i=1,\ldots,n$ placed at the centre of $\bm{y}_p$ is parametrised by a neural  network $f_\theta$ which, at inference time, operates in a sliding-window fashion, taking as inputs patches of fixed size. An example of the proposed map estimation can be appreciated in Fig. \ref{fig:pipeline}.\\
Noise model selection is performed using a standard binary classification network $g_\psi$ able to distinguish between Gaussian and Poisson noise. Several numerical results are reported showing visual and statistical improvements w.r.t.~non-adaptive regularisation models. 
\section{Patch-based learning of regularisation maps} 
We consider in the following an adaptive approach that assigns to each pixel $i\in\left\{1,\ldots n\right\}$ of a noisy image $\bm{y}$ corrupted by Gaussian/Poisson noise a regularisation parameter $\mu_i$ computed by means of a suitable training of a deep neural network leveraging information on a small patch centered in the pixel itself. More precisely, given a patch $\bm{y}_{p_i}$ extracted by $\bm{y}$ around $i\in\left\{1,\ldots n\right\}$, we consider the following patch-wise version of problem \eqref{eq:invariant formulation} suited to deal with Gaussian/Poisson which reads:
\begin{equation} \label{eq:patch_wise_mu}
\argmin_{\bm{x}}~ \mu_i\left(\delta \mathcal{D}_G(\bm{x};\bm{y}_{p_i}) + (1-\delta)\mathcal{D}_P(\bm{x};\bm{y}_{p_i}) \right)+ \text{TV}_\epsilon(\bm{x}),\end{equation}
where $\delta\in\left\{0,1\right\}$ identifies (upon suitable training) whether Gaussian/Poisson noise is observed in the image, so that a tailored data term is used for reconstruction. The parameter $\mu_i$ is the one we would like to estimate to construct the desired regularisation map.
We now parametrise the map $\bm{y}_{p_i}\mapsto \mu_i$  using a deep neural network $f_{\bm{\theta}}$. In the following, we describe the supervised-learning strategy which, given a dataset of clean/noisy natural image patches $\left\{(\tilde{\bm{x}}_p^k, \bm{y}^k_p)\right\}_{k=1}^K$ learn such map. At inference time, note that this map will be applied in a sliding-window fashion, see Fig. \ref{fig:pipeline}, to an unobserved image $\bm{y}$ and used for solving:
\begin{equation} 
\label{eq:whole_image_mu}
\argmin_{\bm{x}}~ \sum_{i=1}^n f_{\bm{\theta}}(\bm{y}_{p_i})\left(\delta d_G(x_i;y_i) + (1-\delta)d_P(x_i;y_i) \right)+ \text{TV}_\epsilon(\bm{x}),
\end{equation}
where thus an adaptive weighting of the data-term against the TV regularisation is computed at each image pixel by considering patches around it.
\subsection{Image patch dataset \& network architecture}
As a training dataset, we considered $300$ images extracted from the popular Berkeley greyscale image dataset (BSD300), covering a diverse range of subjects (from faces, to natural landscapes etc). We preliminary split the dataset into 200/50/50 images for training/validation and testing. 
Overlapping patches of size $32\times32$ were extracted, with a sliding step set to 16 pixels, either to the right or downward, thus letting the patches to overlap by 25\% or 50\%. 
From each extracted patch, several noisy patches generated as different realisations of fixed Gaussian/Poisson noise distributions were generated. 
To create labels, i.e.~reference values $\tilde{\mu}_i$ to be used for training, problem \eqref{eq:patch_wise_mu} was solved repeatedly (see Section \ref{sec:optim}).  
To facilitate the learning process, the outlier values that did not fall in the range $(0, 1.5\times\text{IQR}]$ were removed. The overall samples in the dataset for Poisson denoising are 144,870, for Gaussian denoising are 165,300.
Two separate neural networks were trained for Gaussian and Poisson denoising. Both architectures have approximately $2.8 \times 10^6$ model parameters, as detailed in Table \ref{tab:nn_architecture2}.  Its layer architecture is based on the one of the well-known AlexNet, with modifications aimed at reducing the total number of parameters, so as to mitigate overfitting (via dropout layers). As a training loss the mean squared error (MSE) between the predicted $\mu$ and the reference $\tilde{\mu}$ was used. Adam optimizer was used with an initial learning rate of $10^{-5}$ and an exponential decay rate of factor $\gamma = 0.8$, along with a weight decay of parameter $10^{-4}$. Training and validation were performed over 30 epochs with a batch size of 256.

\begin{table}[h!]
\centering
\begin{tabular}{@{}lcc@{}}
\toprule
\textbf{Layer Type}       & \textbf{Output Shape}        & \textbf{Kernel / Units}                  \\ \midrule
Input Layer               & $(1, 32, 32)$          & -                                 \\ \midrule
Conv2D + BatchNorm + ReLU        & ($64,32,32$)     & $5 \times 5, 64$                              \\ \midrule
MaxPool2D                 & $(64,16,16)$     & $2 \times 2$                                        \\ \midrule
Conv2D + BatchNorm + ReLU       & $(128,16,16)$    & $5 \times 5, 128$                             \\ \midrule
MaxPool2D                 & $(128,8,8)$    & $2 \times 2$                                         \\ \midrule
Conv2D + BatchNorm + ReLU       & $(256,8,8)$    & $3 \times 3$, 256                            \\ \midrule
MaxPool2D                 & $(256,4,4)$      & $2 \times 2$                                       \\ \midrule
Conv2D + BatchNorm + ReLU       & $(512,4,4)$      & $3 \times 3$, 512                              \\ \midrule
MaxPool2D                 & $(512,2,2)$      & $2 \times 2$                                         \\ \midrule
Flatten + BatchNorm1D                  & $2048$                       & -                                            \\ \midrule
Fully Connected + ReLU     & $512$                        & 512                                     \\ \midrule
Dropout (0.25)                 & $512$                        & -                                              \\ \midrule
Fully Connected + ReLU     & $128$                        & 128                                     \\ \midrule
Dropout (0.25)                 & $128$                        & -                                              \\ \midrule
Fully Connected           & $1$                          & 1                                                \\ \bottomrule
\end{tabular}
\caption{Details of the denoising architecture.}
\label{tab:nn_architecture2}
\end{table}

\subsection{Solving the TV problem and computing labels}  \label{sec:optim}

To create labels, i.e.~reference values $\tilde{\mu}_i$ to be used for training, problem \eqref{eq:patch_wise_mu} was solved repeatedly  
for each training patch $\bm{y}_{p}$ being a noisy version of $\tilde{\bm{x}}_p$. More precisely, an optimal parameter $\tilde{\mu}$ and the corresponding solution $\bm{x}_{\tilde{\mu}}$ 
was computed by optimising $\text{SSIM}(\tilde{\bm{x}},\bm{x}_{\mu})$ for different values of $\mu$ chosen within $[\mu_{min}, \mu_{max}]$.

To reduce the computational costs, a golden-section algorithm was employed.
A good range for the parameters was experimentally determined to be $[\mu_{min},\mu_{max}]= [0.01,240]$.
In the case of Gaussian noise, problem \eqref{eq:patch_wise_mu} was solved using accelerated gradient descent (AGD) with fixed step-size. For Poisson denoising, a further step  projecting the solution within the interval $[0,1]$ at each iteration was employed in combination with an adaptive (non-monotone) backtracking strategy \cite{calatroni2019backtracking}. This guarantees faster convergence in comparison with fixed step-size approaches 
which can be quite inefficient in the case of Poisson noise due to the huge Lipschitz constant of the gradient of the KL data term \cite{harmany2011spiral}, which 
leads to small step sizes and slow convergence.

\subsection{Noise model selection: architecture}

For choosing the appropriate data term (and, consequently, the reconstruction pipeline/algorithm) for computing labels, a binary noise classification task (Gaussian VS. Poisson) was performed. For that, patches of size $64\times64$ were used. In comparison to the denoising setting, larger patch size allows here for richer contextual information for distinguishing between noise types. For training the classification network, each clean patch of the dataset was injected with noise of Gaussian/Poisson distribution with varying intensity levels with a total number of 37,800 samples. Specifically, noise intensities were set to $\sigma^2 = \{0.01,0.02,0.03\}$ for Gaussian noise and $\alpha = \{45,30,10\}$ for Poisson noise.
While structurally similar to the architecture detailed in Table~\ref{tab:nn_architecture2}, the classification network differs in some key aspects: it applies higher channel counts in earlier layers (i.e. bigger patches), it lacks a MaxPool2D layer and it ends with a softmax output layer for binary classification. The increased number of channels thus results in a greater number of units in the dense layers so that
the total number of parameters employed is approximately $1.7 \times 10^7$. Training was performed by minimising the cross-entropy loss between predicted and ground-truth classes. As far as the optimization is concerned, the same  setting as the one described for training the parameter estimation network was employed.

\section{Numerical results}

\begin{figure}
    \centering
    \includegraphics[width=1\linewidth]{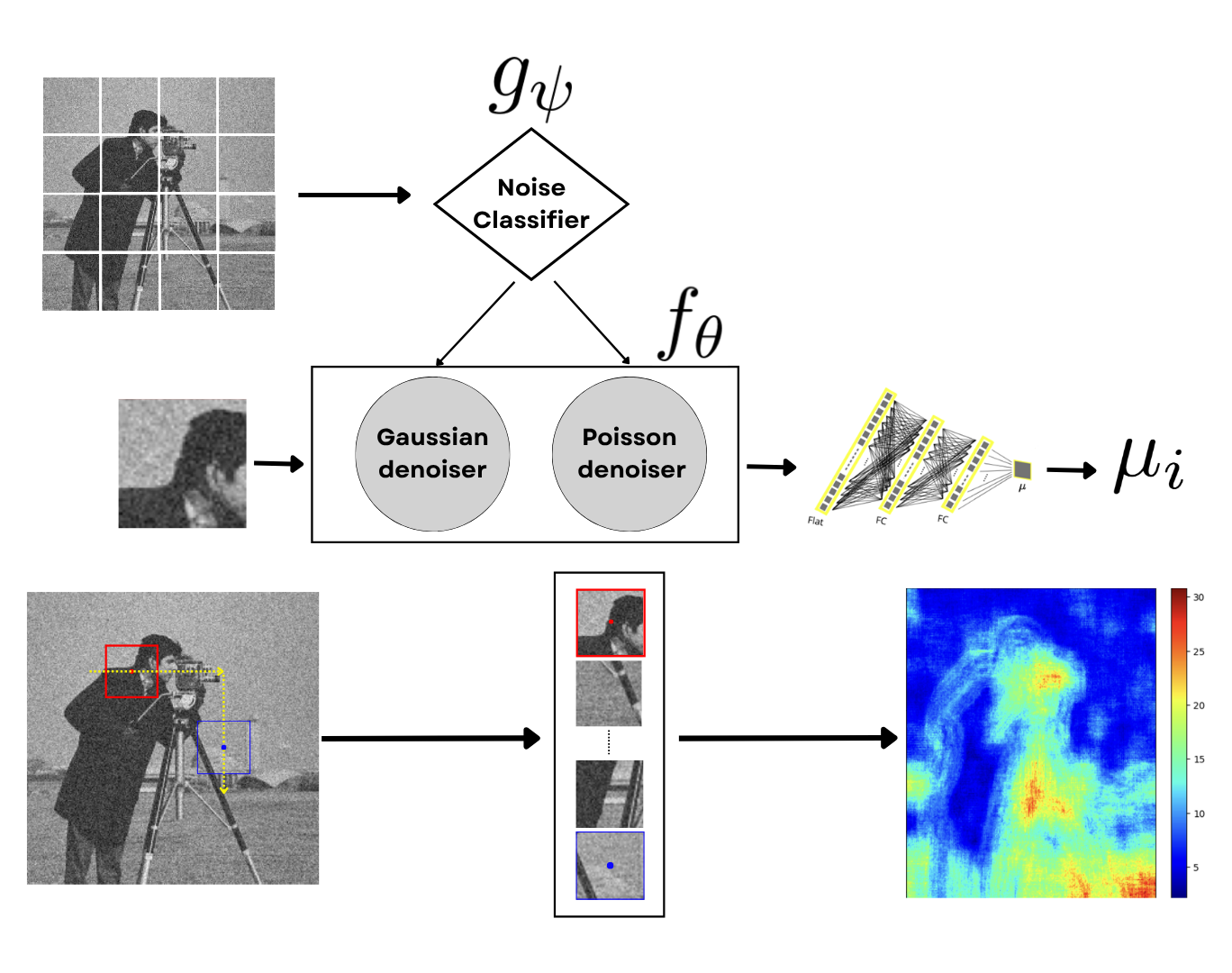}
    \caption{The input noisy image is taken as input by a noise classifier to identify the noise type. Based on the classification result, the corresponding denoising routine is used to estimate a pixel-wise regularisation map in a sliding-window fashion.}
    \label{fig:pipeline}
\end{figure}

The proposed approach is thus divided into two sequential steps: first the noise classifier network is applied to compute whether Gaussian ($\delta=1$) or Poisson ($\delta=0$) noise is observed in the image at hand. Then such parameter is plugged into \eqref{eq:whole_image_mu}, so that the trained network $f_{\bm{\theta}}$ computing pixel-wise the parameter map $\bm{\mu}$ is applied in a sliding-window fashion. A comprehensive overview of our proposed pipeline is shown in Fig.~\ref{fig:pipeline}.
In the following we report the results obtained in correspondence of a Gaussian denoising task with fixed variance $\sigma^2 = 0.01$ and a Poisson denoising task with fixed parameter $\alpha = 30$ after noise model selection is performed.

\subsection{Noise model selection: performance}
The accuracy on the classification task is measured using the following metric
\begin{equation} \label{eq:accuracy}
    \text{NN fitness = }\frac{\text{Number of correctly classified samples}}{\text{Number of samples}}.
\end{equation}
During training the accuracy obtained on the test set is $\text{NN fitness}=0.97$, showing good performance. At inference time, the average accuracy is evaluated over 50 test images, each injected with different noise intensities. Taking as accuracy measure a similar definition to \eqref{eq:accuracy} but measuring the ratio between correctly classified and total number of patches, we obtained  
 the results in Table \ref{tab:noise_accuracy}, covering a range from low noise to higher noise levels.
\begin{table}[h]
    \centering
    \begin{tabular}{l c}
        \toprule
        Noise Type & Avg accuracy (\%) \\
        \midrule
        Gaussian ($\sigma^2 = 0.01$) & 0.67 \\
        Gaussian ($\sigma^2 = 0.015$) & 0.90 \\
        Gaussian ($\sigma^2 = 0.02$) & 0.99 \\
        Gaussian ($\sigma^2 = 0.03$) & 1.0 \\
        Poisson ($\alpha = 60$) & 0.98 \\
        Poisson ($\alpha = 45$) & 0.99 \\
        Poisson ($\alpha = 30$) & 0.99 \\
        Poisson ($\alpha = 15$) & 0.98 \\
        \bottomrule
    \end{tabular}
    \caption{Average accuracy for different types of noise.}
    \label{tab:noise_accuracy}
\end{table}

\subsection{Parameter map estimation: performance}
The fitness of the network on the parameter estimation task is measured using the $R^2$ score:
\begin{equation}
    R^2 := 1 - \frac{\sum_{k=1}^K (\tilde{\mu}_k - \mu_k)}{\sum_{i=1}^K (\tilde{\mu}_k - \mathbbm{E}(\tilde{\mu}_k) )},
    \label{eq:R2_error}
\end{equation}
where $K$ is the the number of test samples, $\tilde{\mu}_k$ are the labels while $\mu_k = f_{\bm{\theta}}({\tilde{\bm{y}}^k_p})$ is the output of the model while tested on the $k$-th patch . \\
The $R^2$ score for Gaussian denoising is 0.67, while for Poisson denoising it is 0.50. Such lower performance could be due to the fact that the Poisson case is more challenging to be dealt with. Although $R^2$ scores are far from values close to 1, our method still achieves good quality reconstruction, as the key factor to consider is the ability to distinguish between regions with varying texture rather than absolute accuracy. \\ 
To validate the performance of the proposed approach, comparisons were performed on a set of test images between the space-variant method \eqref{eq:whole_image_mu} exploiting locally adaptive regularisation, and a scalar approach where a single parameter $\mu_i \equiv \mu$ is computed—via SSIM maximisation (see Sec.~2B)—on the entire image. The two approaches are compared in terms of computational time, PSNR and SSIM.

\subsection{Gaussian denoising $\sigma^2 = 0.01$}

\begin{figure}[h]
    \centering
    \begin{subfigure}[b]{0.51\linewidth}
        \centering
        \includegraphics[width=\linewidth]{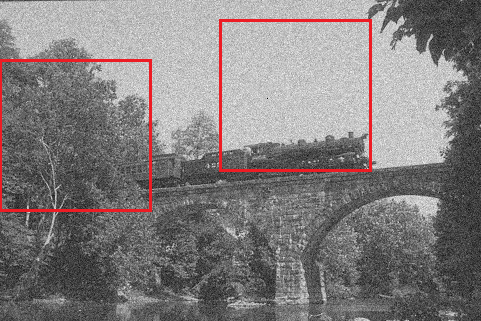}
        \caption{$\bm{y}$, SSIM : 0.4926}
    \end{subfigure}
    \begin{subfigure}[b]{0.47\linewidth}
        \centering
        \includegraphics[width=\linewidth]{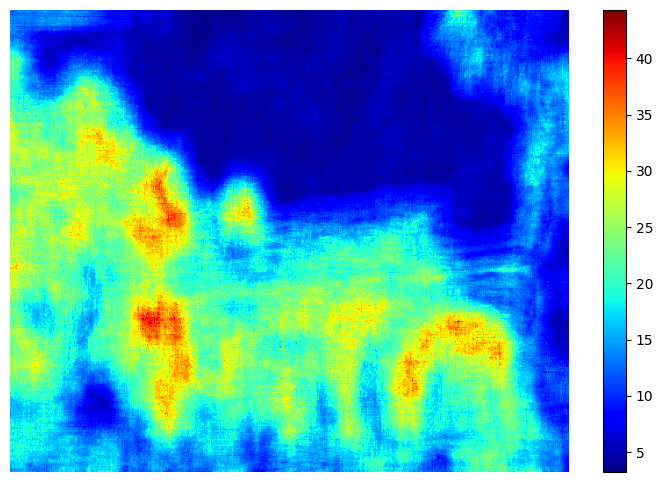}
        \caption{$\bm{\mu}_{\text{map}}$}
    \end{subfigure}
    \begin{subfigure}[b]{0.49\linewidth}
        \centering
        \includegraphics[width=\linewidth]{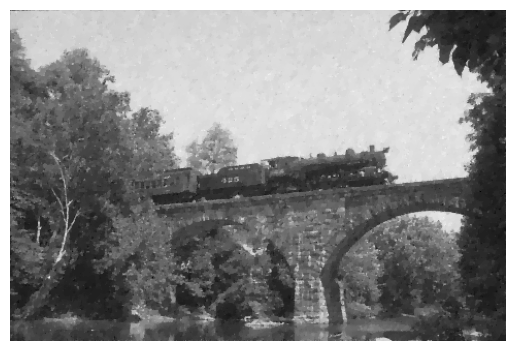}
        \caption{$\bm{x}_{\tilde{\mu}}$, SSIM : 0.7628}
    \end{subfigure}
    \begin{subfigure}[b]{0.49\linewidth}
        \centering
        \includegraphics[width=\linewidth]{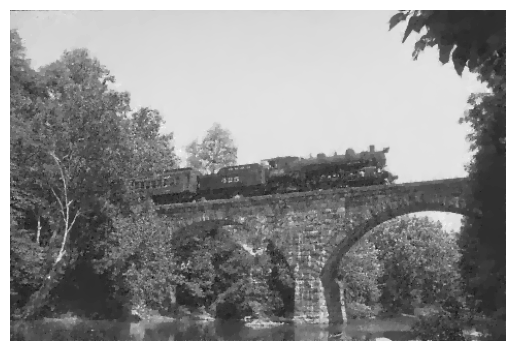}
        \caption{$\bm{x}_{\bm{\mu}_{\text{map}}}$, SSIM : 0.8384}
    \end{subfigure}

    \begin{subfigure}[b]{0.32\linewidth}
        \centering
        \includegraphics[width=\linewidth]{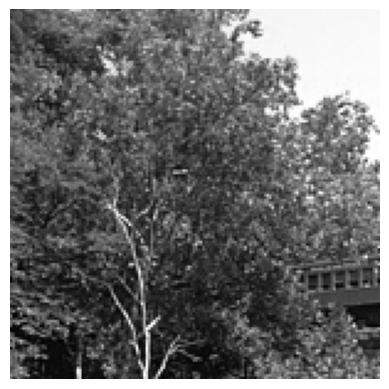}
        \caption{$(\tilde{{\bm{x}}}_p)^1$}
    \end{subfigure}
    \begin{subfigure}[b]{0.32\linewidth}
        \centering
        \includegraphics[width=\linewidth]{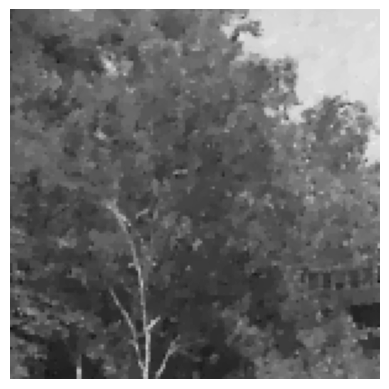}
        \caption{$\bm{x}_{\tilde{\mu}}$, SSIM : 0.685}
    \end{subfigure}
    \begin{subfigure}[b]{0.32\linewidth}
        \centering
        \includegraphics[width=\linewidth]{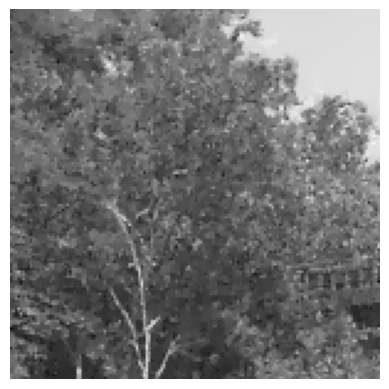}
        \caption{$\bm{x}_{\bm{\mu}_{\text{map}}}$,SSIM:0.784}
    \end{subfigure}

    \begin{subfigure}[b]{0.32\linewidth}
        \centering
        \includegraphics[width=\linewidth]{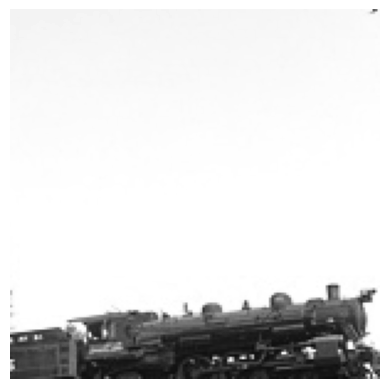}
        \caption{$(\tilde{\bm{x}}_p)^2$}
    \end{subfigure}
    \begin{subfigure}[b]{0.32\linewidth}
        \centering
        \includegraphics[width=\linewidth]{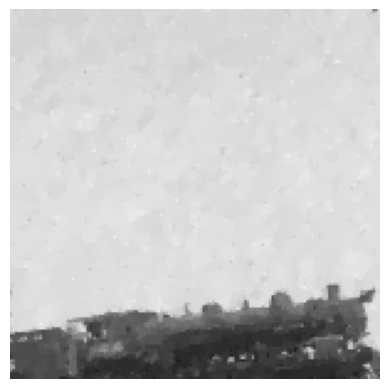}
        \caption{$\bm{x}_{\tilde{\mu}}$, SSIM : 0.848}
    \end{subfigure}
    \begin{subfigure}[b]{0.32\linewidth}
        \centering
        \includegraphics[width=\linewidth]{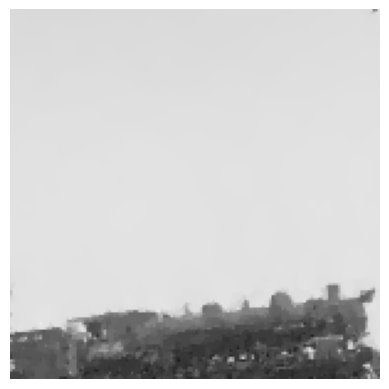}
        \caption{$\bm{x}_{\bm{\mu}_{\text{map}}}$,SSIM: 0.960}
    \end{subfigure}

    \caption{Gaussian denoising $\sigma^2 = 0.01$ on a test image.}
    \label{fig:Gaussian_denoising}
\end{figure}

We report in the following some results obtained in correspondence of a noisy image $\bm{y}$ corrupted with Gaussian noise with zero mean and variance $\sigma^2=0.01$ and show improvements w.r.t.~the scalar approach, see Fig.~\eqref{fig:Gaussian_denoising}. Improvements upon SSIM and PSNR metrics are visible compared to the scalar approach, showing, in particular texture enhancement and stronger smoothing in homogeneous areas. 
We tested the approach over a dataset of $50$ validation images, observing   an average SSIM improvement of 0.017 and PSNR improvement of 0.35 w.r.t.~the scalar approach. 
In terms of computational times, inference required on average $\sim$ 187 seconds of computational time, in comparison of $\sim$ 183 seconds required for computing the solution of the scalar approach with, alongside, optimal scalar parameter selection via golden section which requires direct comparison with the ground truth image and is unfeasible in case of the proposed pixel-dependent parameter map computation.

\subsection{Poisson denoising $\alpha = 30$}

\begin{figure}[h]
    \centering
    \begin{subfigure}[b]{0.505\linewidth}
        \centering
        \includegraphics[width=\linewidth]{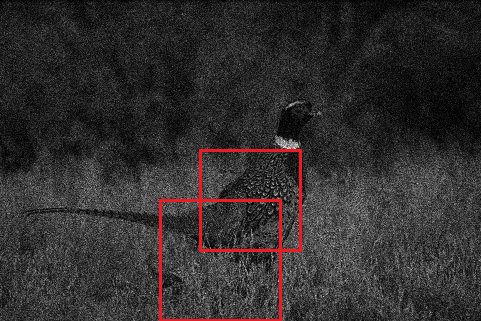}
        \caption{$\bm{y}$, SSIM : 0.344}
    \end{subfigure}
    \begin{subfigure}[b]{0.475\linewidth}
        \centering
        \includegraphics[width=\linewidth]{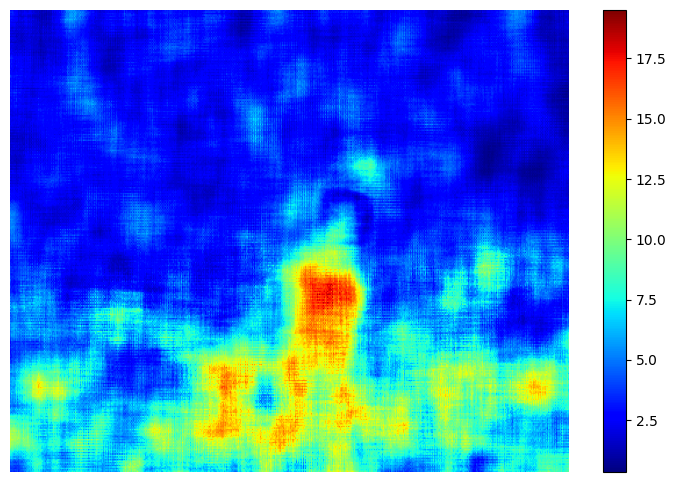}
        \caption{$\bm{\mu}_{\text{map}}$}
    \end{subfigure}

    \begin{subfigure}[b]{0.49\linewidth}
        \centering
        \includegraphics[width=\linewidth]{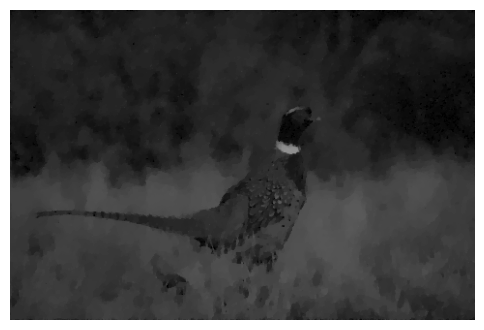}
        \caption{$\bm{x}_{\tilde{\mu}}$, SSIM : 0.683}
    \end{subfigure}
    \begin{subfigure}[b]{0.49\linewidth}
        \centering
        \includegraphics[width=\linewidth]{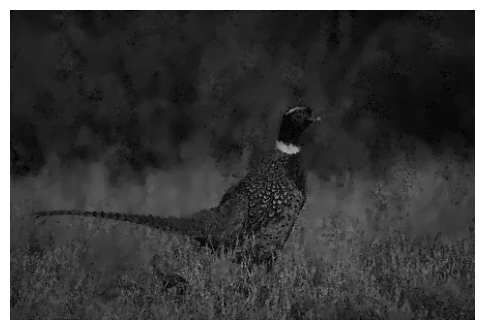}
        \caption{$\bm{x}_{\bm{\mu}_{\text{map}}}$,SSIM : 0.740}
    \end{subfigure}

    \begin{subfigure}[b]{0.32\linewidth}
        \centering
        \includegraphics[width=\linewidth]{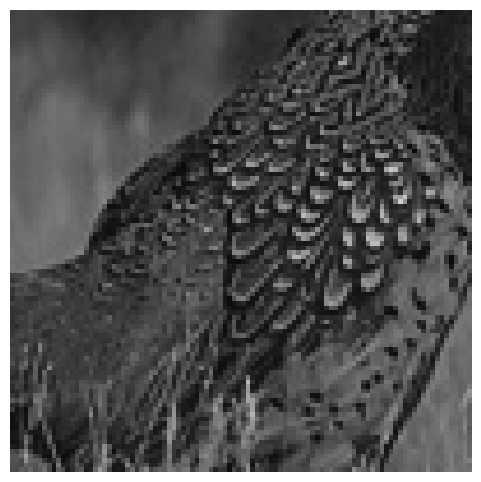}
        \caption{$(\bm{x}_p)^1$}
    \end{subfigure}
    \begin{subfigure}[b]{0.32\linewidth}
        \centering
        \includegraphics[width=\linewidth]{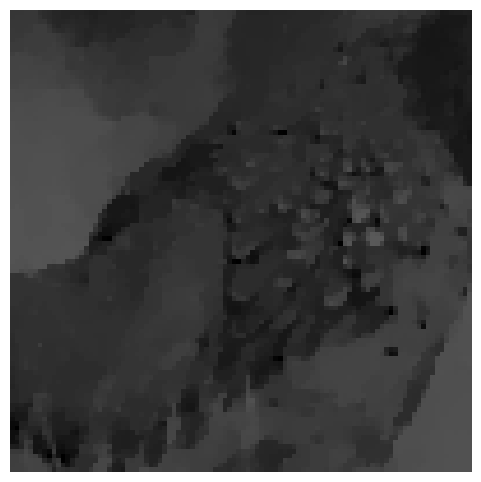}
        \caption{$\bm{x}_{\tilde{\mu}}$,SSIM: 0.479}
    \end{subfigure}
    \begin{subfigure}[b]{0.32\linewidth}
        \centering
        \includegraphics[width=\linewidth]{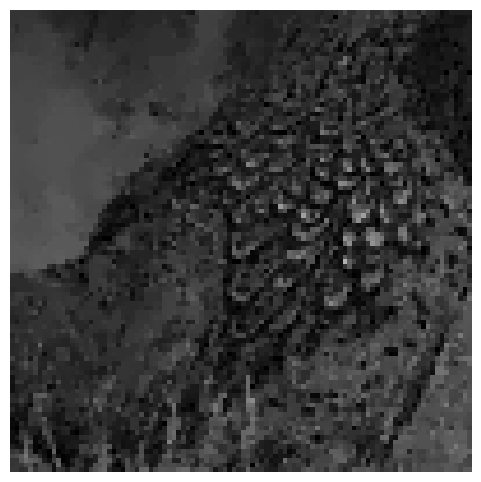}
        \caption{$\bm{x}_{\bm{\mu}_{\text{map}}}$,SSIM:0.707}
    \end{subfigure}

    \begin{subfigure}[b]{0.32\linewidth}
        \centering
        \includegraphics[width=\linewidth]{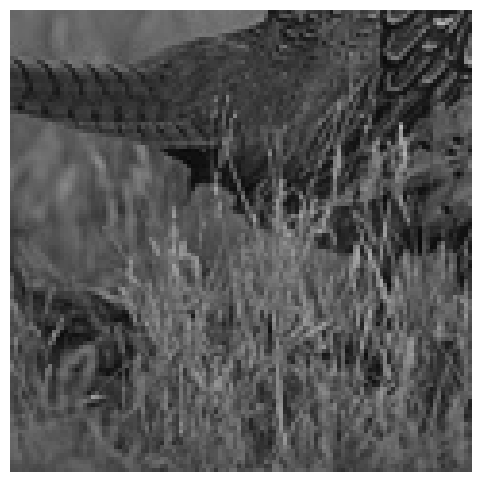}
        \caption{$(\bm{x}_p)^2$}
    \end{subfigure}
    \begin{subfigure}[b]{0.32\linewidth}
        \centering
        \includegraphics[width=\linewidth]{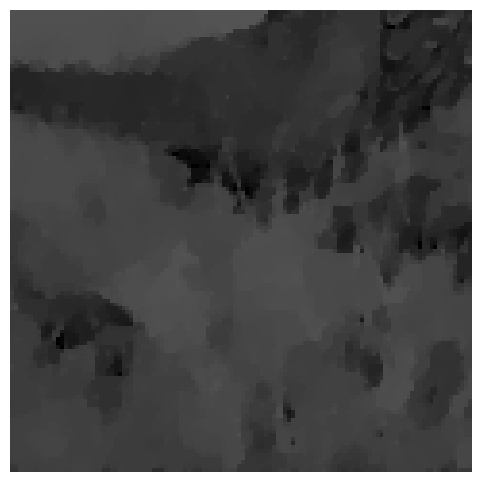}
        \caption{$\bm{x}_{\tilde{\mu}}$,SSIM: 0.413}
    \end{subfigure}
    \begin{subfigure}[b]{0.32\linewidth}
        \centering
        \includegraphics[width=\linewidth]{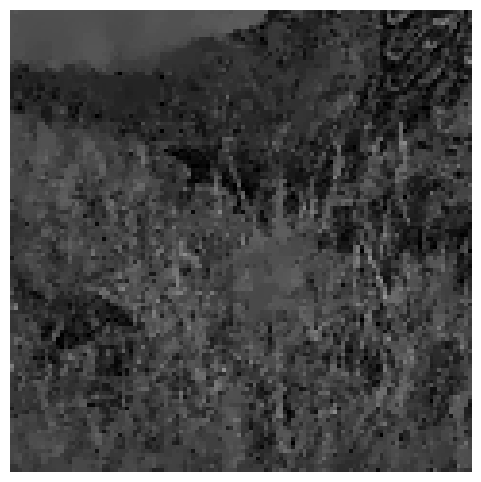}
        \caption{$\bm{x}_{\bm{\mu}_{\text{map}}}$,SSIM: 0.653}
    \end{subfigure}

    \caption{Poisson denoising $\alpha = 30$ on a test image.}
    \label{fig:Poisson_denoising}
\end{figure}

We performed similar tests in correspondence with a Poisson denoising task for images corrupted with a noise level of $\alpha=30$. Fig.~\ref{fig:Poisson_denoising} shows the results obtained in correspondence with an heterogeneous image containing both geometrical and texture areas.
For this task, our method method resulted in an average SSIM improvement  of 0.02 and an average PSNR improvement of 0.02 with respect to the scalar approach.
In terms of computational times, inference time to perform the adaptive parameter map for our approach required $\sim$ 187 seconds, in comparison with computation of the optimal parameter for the scalar approach for the scalar case which required $\sim$ 200 seconds.

\section{Conclusions}

We proposed a patch-wise deep-learning approach for estimating adaptive (i.e., pixel-dependent) regularisation parameter maps for (smoothed) TV image denoising in the presence of either Gaussian or Poisson noise which is detected about preliminary model selection.
Spatial adaptivity combined with tailored noise-model selection allows the model to better handle heterogeneous image structures (geometry VS. texture), thus improving global reconstruction quality while mitigating (over/under)-smoothing effects observed in scalar parameter selection,  in terms of both SSIM and PSNR, while maintaining a reasonable computational cost. Extensions of this work shall address the actual inverse problem setting where a forward operator (such as a convolution or a MRI one as in \cite{Kofler2023}) is incorporated in the modelling.

\bibliography{bibliography}
\bibliographystyle{ieeetr}


\end{document}